\documentclass[journal]{IEEEtran}

\usepackage{romannum}
\usepackage[utf8]{inputenc} % Input encoding
\usepackage[T1]{fontenc}    % Font encoding
\usepackage[english]{babel} % Language hyphenation and typographical rules
\usepackage{lettrine}
\usepackage{comment}

\usepackage{amsmath, amssymb, amsfonts}
\usepackage{mathtools}
\usepackage{braket}

\usepackage{graphicx}
\usepackage{float, stfloats}
\usepackage{caption}
\usepackage{subcaption}
\usepackage{cuted}
\usepackage{color,colortbl}
\definecolor{lightgray}{gray}{0.9}
\definecolor{headerblue}{RGB}{64, 128, 255}
\usepackage{newunicodechar}
\newunicodechar{☆}{\ensuremath{\star}}
\usepackage{siunitx}    % for \si and consistent units

\usepackage{array}
\usepackage{booktabs}
\usepackage{multirow}
\usepackage{tabularx}
\usepackage{booktabs}
\usepackage{caption}
\usepackage{longtable}
\usepackage{ragged2e}
\usepackage{adjustbox}

\usepackage{cite}

\usepackage{url}
\usepackage{hyperref}
\hypersetup{
    colorlinks=true,
    linkcolor=black,
    citecolor=black,
    urlcolor=blue,
}
\usepackage{orcidlink}

\usepackage{xcolor}
\usepackage{algorithm}
\usepackage{algpseudocode}
\usepackage{tikz}
\usetikzlibrary{arrows.meta, positioning}

\title{Quantum Repeater Chains via Cavity–Magnon for Scalable Quantum Networks}

\author{
    \IEEEauthorblockN{Mughees Ahmed Khan   \IEEEauthorrefmark{1}\orcidlink{0009-0009-4214-7637}}
    \IEEEauthorblockN{Syed Shahmir  \IEEEauthorrefmark{2}\orcidlink{0009-0001-1634-2713}}
    \IEEEauthorblockN{M. Talha Rahim \IEEEauthorrefmark{1}\orcidlink{0000-0003-1529-928X}}
        \IEEEauthorblockN{Saif Al-Kuwari  \IEEEauthorrefmark{1}\orcidlink{0000-0002-4402-7710}}
    \IEEEauthorblockN{and Tasawar Abbas  \IEEEauthorrefmark{3}\orcidlink{0000-0002-6048-7346}}   
    
    \IEEEauthorblockA{
        \IEEEauthorrefmark{1}Qatar Center for Quantum Computing, College of Science and Engineering, Hamad Bin Khalifa University, Doha, Qatar}\\
    \IEEEauthorblockA{
        \IEEEauthorrefmark{2}Division of Information and Computing Technology, College of Science and Engineering, Hamad Bin Khalifa University, Doha, Qatar}\\
        \IEEEauthorblockA{
        \IEEEauthorrefmark{3}Department of Physics, COMSATS University Islamabad, 45550 Islamabad, Pakistan.
        }
}

\begin{document}

\maketitle

% *** ABSTRACT ***

\begin{abstract}
Scalable quantum networks require quantum repeaters to overcome major challenges such as photon loss and decoherence in long-distance quantum communication. In this paper, we present a cavity-magnon quantum repeater architecture that exploits the frequency tunability and coherence characteristics of magnonic platforms to enable efficient entanglement swapping across multi-hop networks. Through comprehensive numerical simulations with realistic experimental parameters, we analyze system performance across diverse deployment scenarios and network scales, examining both short-range and long-distance implementations. We identify critical factors influencing performance and scalability, demonstrating that cavity-magnon systems represent a viable and promising quantum repeater platform with significant integration advantages over existing quantum memory technologies.

\end{abstract}

\section{Introduction}
%\lettrine[lines=2]{Q}{\textsc{uantum},} repeaters are a crucial component for long-distance quantum communication, mitigating the effects of loss and noise in quantum channels. 

%\lettrine[lines=2]{R}{\textsc{ecent},} progress in quantum communication research triggered hot research in the field, due to its unique features, i.e., the quantum no-cloning theorem \cite{wootters_single_1982}, its superiority in solving certain complex problems \cite{hussain_empirical_2024}, and security against many quantum-based attacks \cite{slutsky_security_1998}. Thus, increasing attention has been paid to the integration of quantum communication into networks as a key resource for a secure Internet \cite{rohde_quantum_2025}. 

Recent progress in quantum communication promises immense potential to transform the way we secure and transfer information as we harness unique quantum mechanical properties, such as entanglement and superposition. The core of the vision of quantum networks and the global quantum internet, which will connect distant quantum processors and local quantum networks, is to enable powerful applications in cryptography, distributed computing, and quantum-enhanced sensing. However, realizing this vision faces fundamental technical challenges \cite{zoller_quantum_2005}. 
%and quantum transducers.

Building quantum repeaters is not an easy feat, as they require quantum memories, which are challenging to build on their own right \cite{briegel_quantum_1998}. Additionally, factors such as photon loss and decoherence when traveling through optical fibers or aerial links greatly influence the evolution of these quantum systems and limit the long-distance transmission of entanglement. These technical issues even influence the isolated environment of the quantum devices containing the quantum state. Thus, preventing expansion to practical networks due to constraints on the number and efficiency of local gate operations. In classical communication networks, such losses are also observed, but are efficiently corrected by amplifying signals using traditional repeaters. This, however, is not possible in quantum networks due to the no-cloning theorem and the Heisenberg uncertainty principle \cite{wootters_single_1982}. Quantum repeaters have emerged as a potential solution \cite{azuma_quantum_2023}. By breaking down the communication path into shorter, more manageable segments, quantum repeaters generate and maintain entanglement within each segment. Through a process called entanglement swapping \cite{hu_progress_2023}, these shorter segments are linked together to extend the quantum states across greater distances. Essentially, quantum repeaters teleport quantum information from one node to another without requiring the physical transmission of photons between the two ends, significantly reducing photon loss and maintaining quantum coherence.

Considerable progress has already been made in developing entangled correlation in various quantum-based setups, such as cavity quantum electrodynamics (Cavity-QED) \cite{van_enk_quantum_2004, ali_teleportation_2022, ali_hyperentanglement_2022}, circuit-quantum electrodynamics (Circuit-QED) \cite{shahmir_multi-party_2023, yang_generating_2013}, nuclear magnetic resonance (NMR) \cite{xin_nuclear_2018}, spintronics \cite{da-xiu_experimental_2004} and photonic setups \cite{andersen_step_2021}, for which many theoretical schemes were proposed and later verified experimentally. In addition to gate implementation, these systems have been explored to be used to develop quantum repeaters \cite{krutyanskiy_telecom-wavelength_2023, kumar_towards_2019}. All these quantum systems have different controls and come with their own constraints. Among them, cavity magnonics holds significant potential because its tunable magnon–photon coupling, long coherence, and straightforward interface with both optical and superconducting circuits make it particularly amenable to hybrid integration \cite{zare_rameshti_cavity_2022}.

Considerable experimental milestones for quantum networks have been achieved over the past few years, with quantum repeater technology rapidly advancing from laboratory prototypes to deployable infrastructure. In 2020, the USTC group in China achieved a significant milestone by extending memory-to-memory entanglement over 12.5 km, achieving city-scale distribution using atomic-ensemble memories \cite{liu_creation_2024}. In 2021, a team at Delft linked three nitrogen-vacancy–center nodes through entanglement swapping over a separation of 1.3 km \cite{ruf_quantum_2021}. Building on this, in 2022, the same group succeeded in teleporting quantum information across non-neighboring nodes \cite{hermans_qubit_2022}, marking a major step forward. In 2023, researchers at the University of Innsbruck demonstrated entanglement swapping over a fiber channel using an inter-telecom wavelength, generating heralded entanglement between two distant parties over 50 km \cite{krutyanskiy_telecom-wavelength_2023}. Furthermore, recent demonstrations on existing metropolitan networks, such as the AWS-Harvard link in Boston (about 35 km), the Argonne 84 km Chicago test loop, and the Qunnect 16 km GothamQ network in New York,  proved the robustness of quantum hardware outside the laboratory. In addition, various national initiatives under the European Union and elsewhere have established testbeds for quantum communication \cite{raymer_us_2019, riedel_europes_2019}. Collectively, these achievements have charted a firm trajectory towards repeater-enabled quantum networks.

Recently, cavity magnonics has emerged as a hybrid architecture for the integration of the cavity-QED or circuit-QED with YIG-magnon quantum systems \cite{zare_rameshti_cavity_2022}. Similarly, the integration of superconducting circuits with YIG magnons has also been studied \cite{dols_magnon-mediated_2024, zhang_tunable_2024}. This allows for various cost-effective engineered systems with better controllability for qubits. On one hand, superconducting qubits are excellent processors and have high-fidelity gates. These qubits have different engineered versions that are prone to different kinds of dissipation, making them impractical for long-term data storage. On the other hand, magnons show promising properties for quantum memories, interaction mediators, and hosting different regimes of coupling, i.e., optomechanical and optical. Furthermore, the integration of this paradigmatic material, such as to operate as Yttrium Iron Garnet (YIG) magnon, gives reduced dissipative magnons, which gives the quantum state a longer coherence time.

The ability to establish remote entanglement between two or more distant quantum objects, magnons in our case, is a fundamental step towards building large-scale quantum networks. These quantum networks will rely on a well-defined physical structure and topology to effectively transfer quantum information, yet they remain constrained by inherent limitations, such as finite entanglement generation rates and limited coherence times \cite{chehimi_scaling_2023, noauthor_practical_2024}. By choosing a hybrid quantum system, which tends to generate and keep these states for a longer period, i.e., better lifetimes of these quantum states. The longer coherence time results in a larger number of computational tasks per quantum state and is also an enabler for communication. This allows one to circumvent many of the limitations of standard quantum systems and avoid bottlenecks faced by many such quantum networks, thus allowing these setups to be tailored to specific needs. This approach is uniquely advantageous in quantum networks, as quantum communication resources can be prepared in advance. 

\subsection{Contributions} 
In this paper, we focus on using a hybrid cavity magnonics structure to generate entanglement over long distances, thereby enabling flexible quantum communication. Our cavity‑magnonic system works as a quantum repeater in which remote entanglement is generated and extended via successive entanglement swapping operations. The setup has been shown to have better coherence time and longer intrinsic state memories as magnonic systems are flux tunable, can be engineered to have lower cooling overhead, and are shown to be compatible with both optical and microwave systems \cite{kounalakis_flux-mediated_2020, wu_remote_2021}. Our proposed setup consists of identical elementary links that connect these individual setups remotely. In our study, we investigate two scenarios related to communication infrastructure. The first scenario involves a compact setup that can be integrated onto a chip, allowing for implementation within the microwave domain. The second scenario addresses larger systems, such as those deployed at a city scale, where microwave-to-optical converters are necessary to bridge the connection with optical communication networks. 
 
We benchmark the proposed cavity–magnon repeater using a composite figure of merit that couples entanglement quality, as captured by concurrence and conditional Bell-state fidelity, with the distribution rate embodied in the heralded success probability, by simulating these metrics under realistic dissipation, fibre attenuation and multiplexing in both chip-scale and metro-scale scenarios. All numerical results use parameter sets drawn directly from state-of-the-art microwave-cavity, optical-conversion and dense-multiplexing experiments, ensuring that our simulations reflect realistic hardware performance while isolating the architecture-level design trade-offs.

\subsection{Organization}
The rest of this paper is organized as follows. Section \ref{cav_mag ent} provides a short review of cavity-magnon interaction using the Lindblad master equation.
In Section \ref{qres}, we define and compute key figures of merit, including concurrence and fidelity, while tracking their time-dependent populations to quantify the quality of the states we generate. In Section \ref{qrepnet}, we present an entanglement-swapping protocol, demonstrating how chaining these swaps can create a scalable cavity-magnonic repeater architecture for distributing heralded Bell pairs over long distances. In Section \ref{resultsNdiscussion}, we present our numerical results and provide a detailed discussion of their implications. Finally, the paper concludes in Section \ref{conclusion}.

\section{Preliminaries}\label{cav_mag ent}
In this section, we review the fundamentals of magnon-photon interactions in open quantum systems, establishing the theoretical foundation for cavity-magnon entanglement generation. We begin by introducing the quantum resource metrics, namely concurrence and fidelity, which serve as benchmarks for evaluating our proposed quantum repeater scheme. We then examine cavity-magnon entanglement generation, beginning with the ideal unitary dynamics and progressing to realistic open-system evolution under environmental noise, with numerical simulations demonstrating high-fidelity Bell state preparation using experimentally relevant parameters.

\subsection{Quantum Resources}\label{qres}
In quantum communication networks, the quantification of non-classical correlations is vital to assess protocol performance and ensure reliable entanglement distribution over the noisy channel. Here, the resource theories provide a formal framework for evaluating these quantum features, offering resource monotones that measure the utility of quantum states under a given set of allowed operations. In this work, in order to evaluate the performance of our proposed entanglement swapping-based quantum repeater protocol in the cavity magnonic architecture,  we focus on two key resource metrics, concurrence and fidelity.  

\paragraph{Concurrence} Concurrence is a widely used monotone to quantify quantum entanglement \cite{wootters_entanglement_1998}, where higher values of concurrence indicate robust quantum entanglement, commonly serving as a critical quantum resource in entanglement-based quantum communication and quantum repeater setups. For a two-qubit density matrix $\rho$, the concurrence $\mathcal{C}(\rho)$ is defined as,
\begin{equation}
    \mathcal{C}(\rho)=max\{0,\lambda_1 - \lambda_2 - \lambda_3 - \lambda_4\}
\end{equation}
where $\lambda_i$ represents the square roots of the eigenvalues of the matrix $\rho (\sigma_y \otimes \sigma_y) \rho^* (\sigma_y \otimes \sigma_y)$, arranged in decreasing order, and $\rho^*$ denotes the complex conjugate of the density matrix. The boundaries of the concurrence intervals represent the separable state \(\mathcal{C}(t) = 0\) and the maximally entangled state \(\mathcal{C}(t) = 1\).  In our analysis, we track concurrence dynamically to investigate how entanglement between magnonic subsystems evolves under various interactions for end-to-end entanglement generation under the losses described in the evolution of interaction and swap operations.

\paragraph{Fidelity} Fidelity is a fundamental metric in quantum information and is used to quantify the similarity between two quantum states. For two density matrices \(\rho\) and \(\sigma\), representing the actual and ideal quantum states, respectively, the fidelity of the states is defined as,
\begin{equation}\label{fids}
    \mathcal{F}(\rho, \sigma) = \left( \mathrm{Tr} \left[ \sqrt{ \sqrt{\rho} \, \sigma \, \sqrt{\rho} } \right] \right)^2
\end{equation}

The fidelity bounded by \(0 \leq \mathcal{F}(\rho,\sigma) \leq 1\) where \(\mathcal{F}=1\) implies that the two density matrices are identical. Fidelity is crucial in assessing the success of quantum operations such as teleportation, entanglement swapping, and error correction. In practical applications, fidelity helps identify decoherence and noise, serving as a guiding metric for optimizing experimental parameters.  In our repeater framework, fidelity is used to evaluate the degree to which the resulting entangled magnonic state approximates the desired Bell state after environmental noise and operational imperfections cumulating at each repeater node.

\subsection{Cavity-magnon Entanglement Generation}
In cavity magnonic systems, the composite system consists of ferrimagnetic materials, such as Yttrium-Iron-Garnet (YIG) spheres embedded in high-quality microwave cavities, which enable the formation of hybridized magnon–photon states through tunable magnetic biasing and offer several advantageous properties for quantum information processing, such as long coherence times and tunability with wide frequency domains along with strong coupling to enable efficient quantum state transduction \cite{wolz_introducing_2020, yuan_quantum_2022}. In our setup, we take a simple case where we use single mode field and a two-level truncated magnon for which the total Hamiltonian is given as \cite{zare_rameshti_cavity_2022},

\begin{align}\label{eq:H_o}
    \mathcal{H}_o = \omega_c c^\dagger c + \omega_m m^\dagger m + g_{mc} (m + m^\dagger)(c + c^\dagger),
\end{align}
Here, the first two terms are the bare energies of the cavity and magnon modes with \(c^\dagger\) \((c)\), \(m^\dagger\) \((m)\) their respective creation (annihilation) operators for the cavity and magnon modes and the last term describes the coherent exchange between the subsystems. The coupling strength is, $g_{mc} = \gamma \sqrt{\hbar \omega_c \mu_0  S/2 V_c}$ with $\gamma$ is the gyromagnetic ratio, $\mu_o$ is the vacuum permeability, 
$V_c$ is the mode volume of the cavity, and 
\(S\) is the total spin of the magnon ensemble. Assuming perfect spatial overlap between the cavity's magnetic field and the magnon mode and working near resonance \(\omega_c \approx \omega_m\) in the weak coupling regime \( g_{mc} \ll \omega_c, \omega_m \) the evolution under the rotating-wave approximation (RWA) gives the effective Hamiltonian as,
\begin{align}\label{eq:H_int}
    \mathcal{H}_{i} = g_{mc} (m^\dagger c + c^\dagger m ).
\end{align}

For our system, we consider the initial state \(\ket{\psi(0)} = \ket{0}_m \ket{1}_c\), where \(m\) and \(c\) denote the magnon and cavity modes, respectively. 
For the ideal case, the system's evolution under \(\mathcal{H}_{i}\) is given by \(\ket{\psi(t)} = \mathcal{U}(t)\ket{\psi(0)}\), where the time-evolution operator is \(\mathcal{U}(t) = \exp(-i \mathcal{H}_i t)\). 
The evolved state at the interaction time \(t = \pi/4g_{mc}\) is then given by:

\begin{align}\label{eq3}
    \ket{\psi(t)} \;=\; \frac{\ket{0_m,1_c} 
-\,i\, \ket{1_m,0_c}}{\sqrt{2}},
\end{align}
Here, $\ket{\psi(t)}$ is a maximally entangled state locally equivalent to the singlet $\ket{\psi^-}$ which can be represented in the density matrix formalism as $ \rho = \ket{\psi(t)}\bra{\psi(t)} $. 
For the non-ideal case corresponding to realistic scenarios, environmental interactions introduce noise effects to the quantum system. 

\begin{figure}
  \centering
  \includegraphics[width=0.45\textwidth]{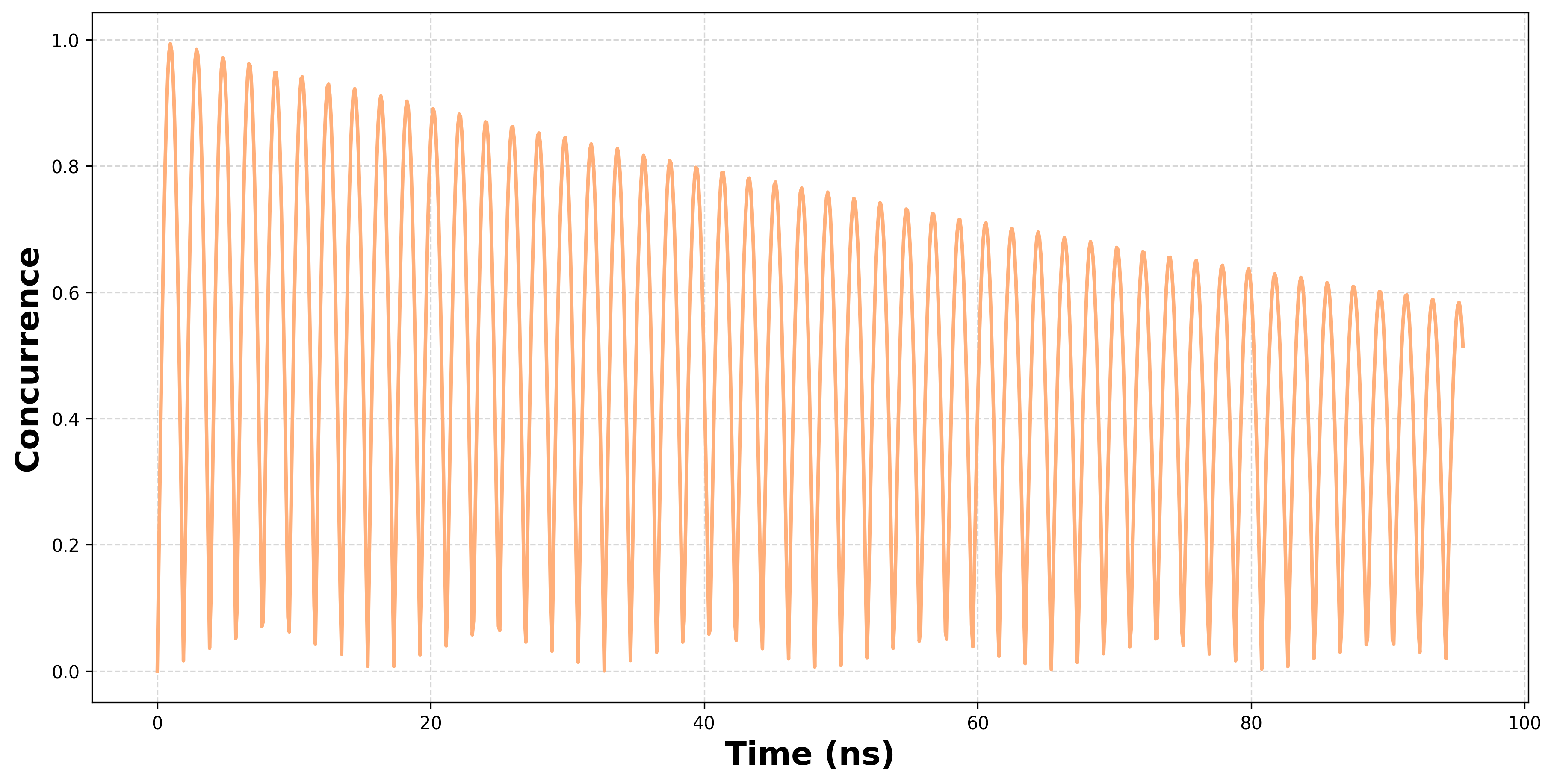}
  \caption{
    Concurrence as a function of time.  
    Simulation parameters: \(\omega_c/2\pi = 10\) GHz, \(\omega_m/2\pi = 10\) GHz, \(g_{mc}/2\pi = 130\) MHz, \(\kappa_d/2\pi = 1\) MHz, \(\gamma_d/2\pi = 0.5\) MHz, \(\kappa_\phi/2\pi = 0.3\) MHz, \(\gamma_\phi/2\pi = 0.3\) MHz.
  }
  \label{fig:cavity_magnon_dynamics}
\end{figure}

\begin{figure}
  \centering
  \includegraphics[width=0.48\textwidth]{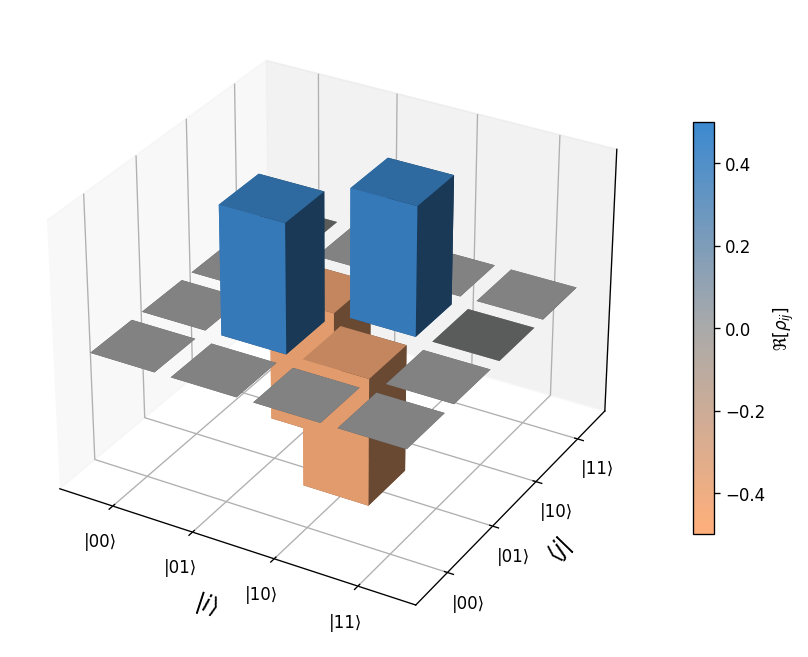}
  \caption{
    The final two-qubit state is represented as a 3D density matrix histogram.  
    Simulation parameters: \(\omega_c/2\pi = 10\) GHz, \(\omega_m/2\pi = 10\) GHz, \(g_{mc}/2\pi = 130\) MHz, \(\kappa_d/2\pi = 1\) MHz, \(\gamma_d/2\pi = 0.5\) MHz, \(\kappa_\phi/2\pi = 0.3\) MHz, \(\gamma_\phi/2\pi = 0.3\) MHz.
  }
  \label{fig:singlet_density_matrix}
\end{figure}

The Lindblad formalism provides a framework for describing the dynamical evolution of such open quantum systems, incorporating dissipation, decoherence, and other non-unitary effects added phenomenologically. Mathematically, this is expressed as
\begin{equation}\label{eq:lindblad}
   \frac{d\rho}{dt} = \frac{-i}{\hbar}[ \hat{\mathcal{H}}, \hat{\rho}] 
+ \mathcal{D}(\hat{\rho}).
\end{equation}
Here, the first term represents the unitary evolution of the system described by the density operator \(\hat{\rho}\) under the Hamiltonian \(\hat{\mathcal{H}}\). The second term describes the dissipative dynamics, which account for energy exchange with the environment. The super-operator \(\mathcal{D}(\hat{\rho})\) is given by \cite{scully_quantum_1997}
\begin{equation}
    \mathcal{D}(\hat{\rho}) = \sum_k \left( \mathcal{L}_k \hat{\rho} \mathcal{L}_k^\dagger 
- \frac{1}{2} \{\mathcal{L}_k^\dagger \mathcal{L}_k, \hat{\rho}\} \right).
\end{equation}
Here, the collapse operators \(\mathcal{L}_k\) characterize the dephasing and dissipation effects within the cavity and magnon modes which are expressed as \(\mathcal{L}_c=\sqrt{\kappa_d}\hat{c}\,\), \(\mathcal{L}_m = \sqrt{\gamma_d}\,\hat{m},\) for decay rates and \(\mathcal{L}_{c_\phi} = \sqrt{\kappa_\phi}\, \hat{c}^\dagger \hat{c}\),
\(\mathcal{L}_{m_\phi} = \sqrt{\gamma_\phi}\, \hat{m}^\dagger \hat{m}.\) for the dephasing noise effects of the cavity and magnon modes, respectively \cite{scully_quantum_1997}. To verify the presence of entanglement and quantify the underlying correlations, we evaluate the system's state over the effective interaction period shown in Fig. (\ref{fig:cavity_magnon_dynamics}). Using the QuTiP library, we reconstruct the final state via its built-in quantum state tomography functions. The resulting density matrix, shown in Fig. (\ref{fig:singlet_density_matrix}), demonstrates the formation of the Bell type state in Eq. \eqref{eq3}, and provides a comprehensive visualization of the system's coherence and population under non-ideal conditions.

In Fig. (\ref{fig:singlet_density_matrix}), the density matrix representation gives the chromatic intensity gradation to quantitatively depict the magnitude of the constituent elements after Bell-State Measurement (BSM). Here, we focus specifically on the strong coupling regime, defined by the condition \(g_{mc} \gg (\kappa+\gamma)/2\) where \(\kappa = \kappa_d + \kappa_{\phi}\) and \(\gamma = \gamma_d + \gamma_{\phi}\) are the total dissipation of the respective subsystems. Furthermore, we take the resonant frequency of the system to be at $\omega_i = 2\pi \times 10 \;\text{GHz}$, $i \in \{ c,m \}$ for both cavity and magnon modes, respectively, with a coupling rate of $g_{mc} = 2\pi \times130 \;\text{MHz}$, giving the maximum concurrence at approximately 0.97. We assume the microwave cavities at low temperatures with quality factor in the range \((10^4-10^6)\) and cavity decay rates of $ \kappa_d =  2\pi \times 1 \;\text{MHz}$. The magnon damping rates, primarily affected by intrinsic Gilbert damping in YIG and material imperfections are taken to be $\gamma_d = 2\pi \times0.5 \;\text{MHz}$, while pure dephasing rates for both cavity and magnon modes were maintained at $ \kappa_\phi = \gamma_\phi = 2\pi \times 0.3 \;\text{MHz}$. Analysis reveals prominent off-diagonal components (magnitude \(\approx \pm 0.47)\) that indicate substantial quantum coherence and non-classical correlations between the bipartite subsystems. This coherence manifests with notable phase symmetry, consistent with theoretical predictions the Bell type state with minimal decoherence effects. The observed matrix structure demonstrates high-fidelity entanglement preparation. 

In our setup, we selected an experimentally relevant time window $t$ using achievable parameters, thus demonstrating the system's practical potential and robustness. In the subsequent section, we utilize this quantum state tomography to reconstruct the density matrix after subsequent entanglement swapping operations to benchmark our quantum repeater architecture.

\section{Quantum Repeater Networks}\label{qrepnet}
\begin{figure}
  \centering
  \includegraphics[scale=0.065]{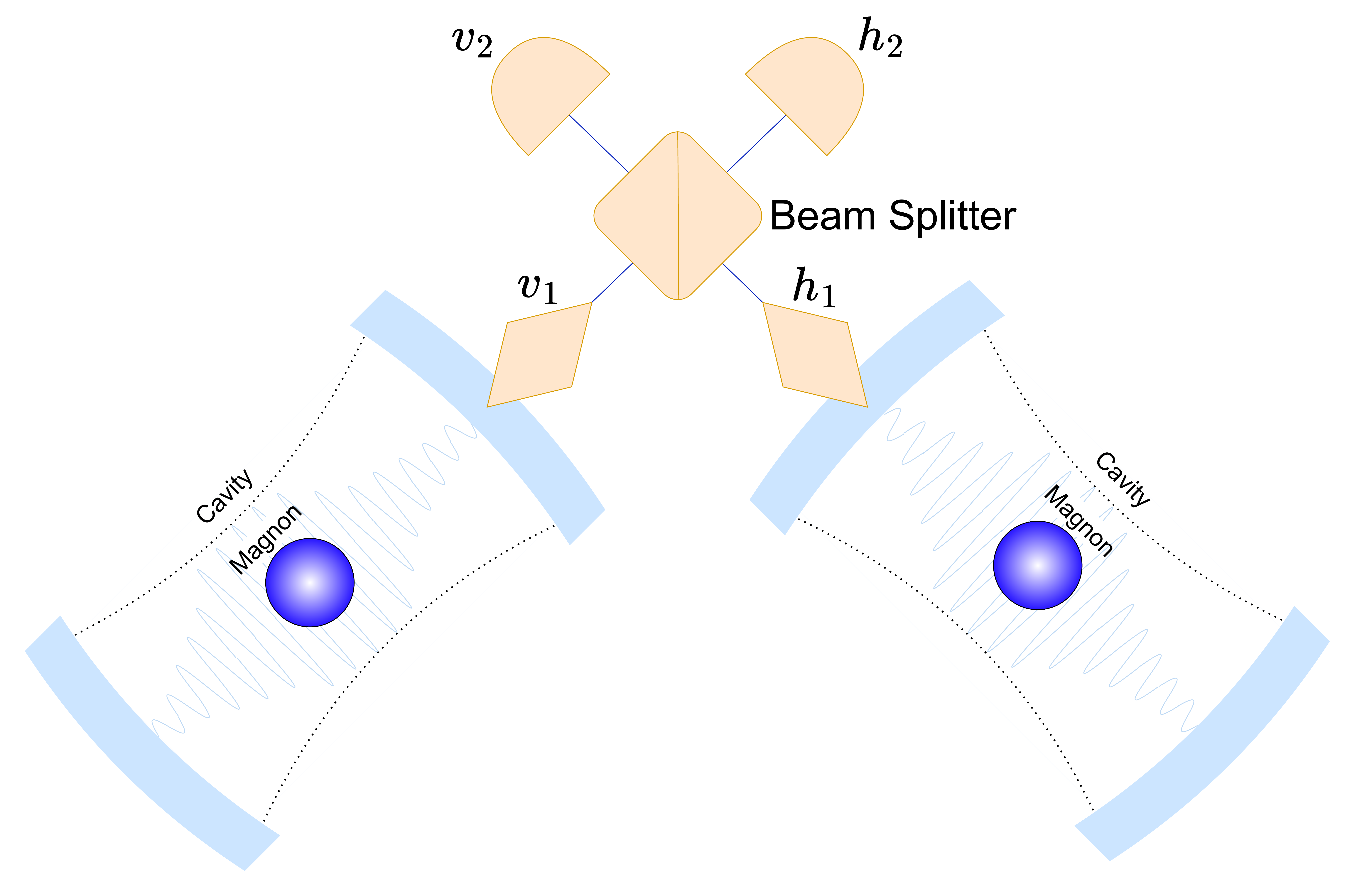}
  \caption{%
    Schematic illustrating the functionality of a beam splitter, along with the corresponding input–output relationships in the cavity magnonic system.
  }
  \label{fig:enswap}
\end{figure}

Quantum repeater networks are based on the fundamental phenomenon of entanglement swapping, a key technique for distributing entanglement between two distant nodes without direct interaction.  These devices are designed to extend the range of quantum communication by overcoming the limitations imposed by signal loss and decoherence in optical fibers. In this section, we consider two identical nodes consisting of cavity-magnon setups as subsystems in a network connected over a fiber channel, as depicted in Fig. (\ref{fig:enswap}). At the outset, each node is prepared in identical entangled states, each being $\rho_{loc} = \ket{\psi^-}\bra{\psi^-}$, where $\ket{\psi^-}$ is the Bell state obtained using Eq. \eqref{eq3}. After the generation of local entanglement between respective cavity-magnons, the optical field is allowed to travel a distance $L$ over the fiber channel before being mixed at the beam splitter by transformation described as,
\begin{equation}\label{intHam}
    \mathcal{H}_{bs} = \beta(a^\dagger b + ab^\dagger)
\end{equation}
where $a^\dagger, a \; (b^\dagger \; b)$ are the Bosonic operators of the respective cavity modes traveled to the beam-splitter and evolve by the unitary operation expressed explicitly as,
\begin{equation}
    \mathcal{U}_{bs} = \left(
\begin{array}{cccc}          
\cos (\beta t) & -i \sin (\beta t)  \\
-i \sin (\beta t) & \cos (\beta t)  \\
\end{array}
\right)
\end{equation} 
This effectively removes the ``which-path'' information, leading to maximal mixing of states. 

\begin{figure}
  \centering
  \includegraphics[width=0.5\textwidth]{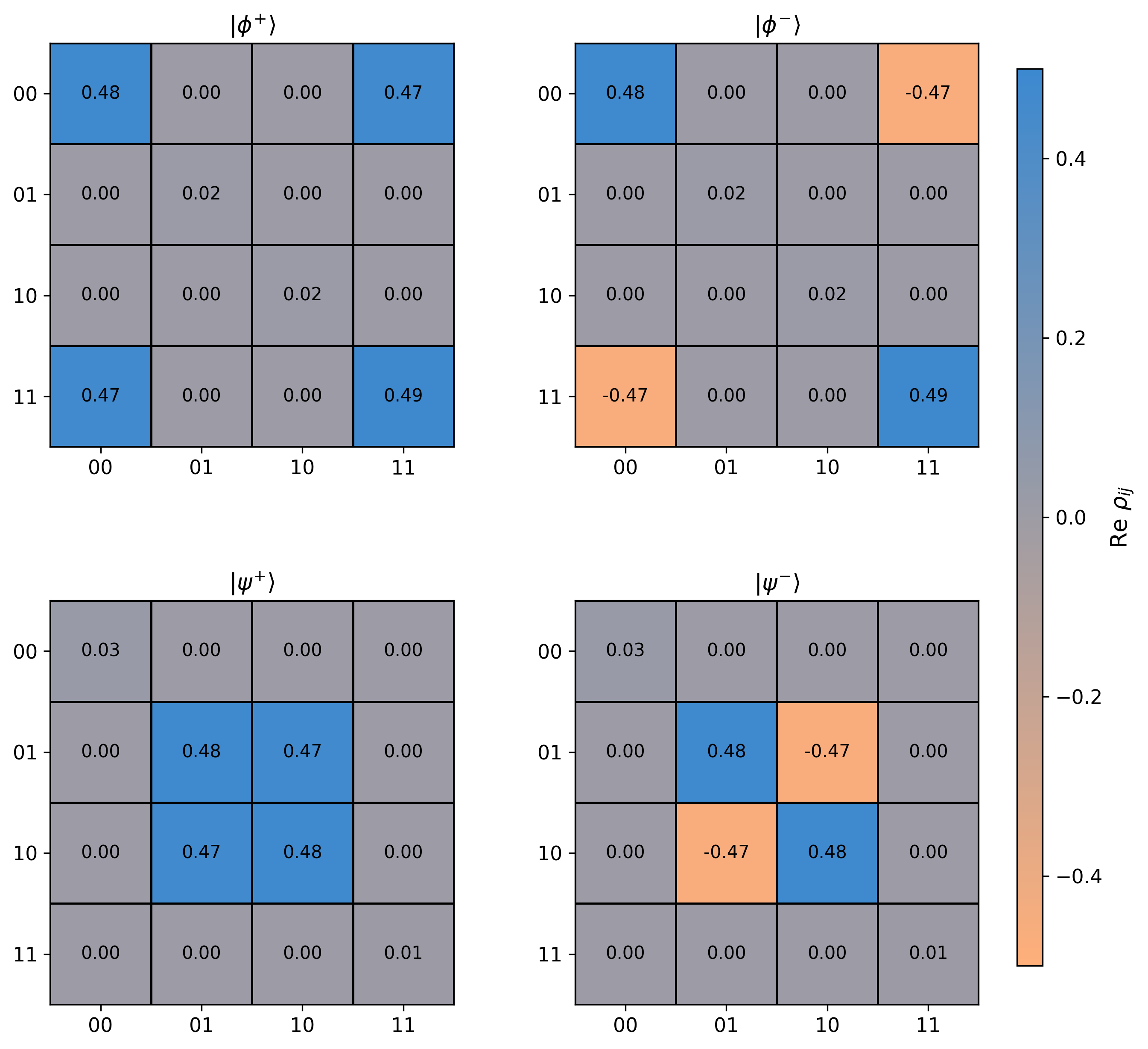}
  \caption{%
    Magnon–magnon interference fringes and reconstructed density matrix after entanglement swapping at a single node.
  }
  \label{fig:singlet}
\end{figure}

In the case of a symmetric beam splitter, we have \(|\cos(\beta t)|^2 = |\sin(\beta t)|^2 = \frac{1}{2}\). In this scenario, the reflected beam experiences a relative \(\pi/2\)-phase shift, while the transmitted beam remains unchanged. This effectively mixes the quantum states within the system. As a result, the input-output relations for the cavity fields can be expressed as,
\begin{equation}
                    \left(
                            \begin{array}{c}          
                                v_2  \\
                                h_2  \\
                            \end{array}
                     \right) = \mathcal{U}_{bs} 
                     \left(
                            \begin{array}{c}          
                                v_1  \\
                                h_1  \\
                            \end{array} 
                    \right)
\end{equation}
Here, \(v_2\) and \(h_2\) are the outputs corresponding to the incoming input photons \(v_1\) and \(h_1\), as illustrated in Fig. (\ref{fig:enswap}). By detecting the output photons at the detectors by BSM, we have the success probability of $50\%$ and decay rate for each photon in the fiber link is $e^{-L/2d}$. Thus, in a typical scenario, the success probability of the BSM operation to generate the entangled state between magnons at end nodes is $\mathcal{P}_{\text{succ}} = e^{-L/d}/2$, where $d=10 \text{km}$. The detection gives us a heralding signal that can determine the outcome state of entanglement from one of the four Bell pairs between the magnons as shown in Fig. (\ref{fig:singlet}).  In case of non-linear Bell state analyzers, this success probability can increase to $100\%$ by incorporating corrections given in Table \ref{bsa}\cite{stephenson_high-rate_2020}. 

\begin{figure}
    \centering
    \includegraphics[scale=0.075]{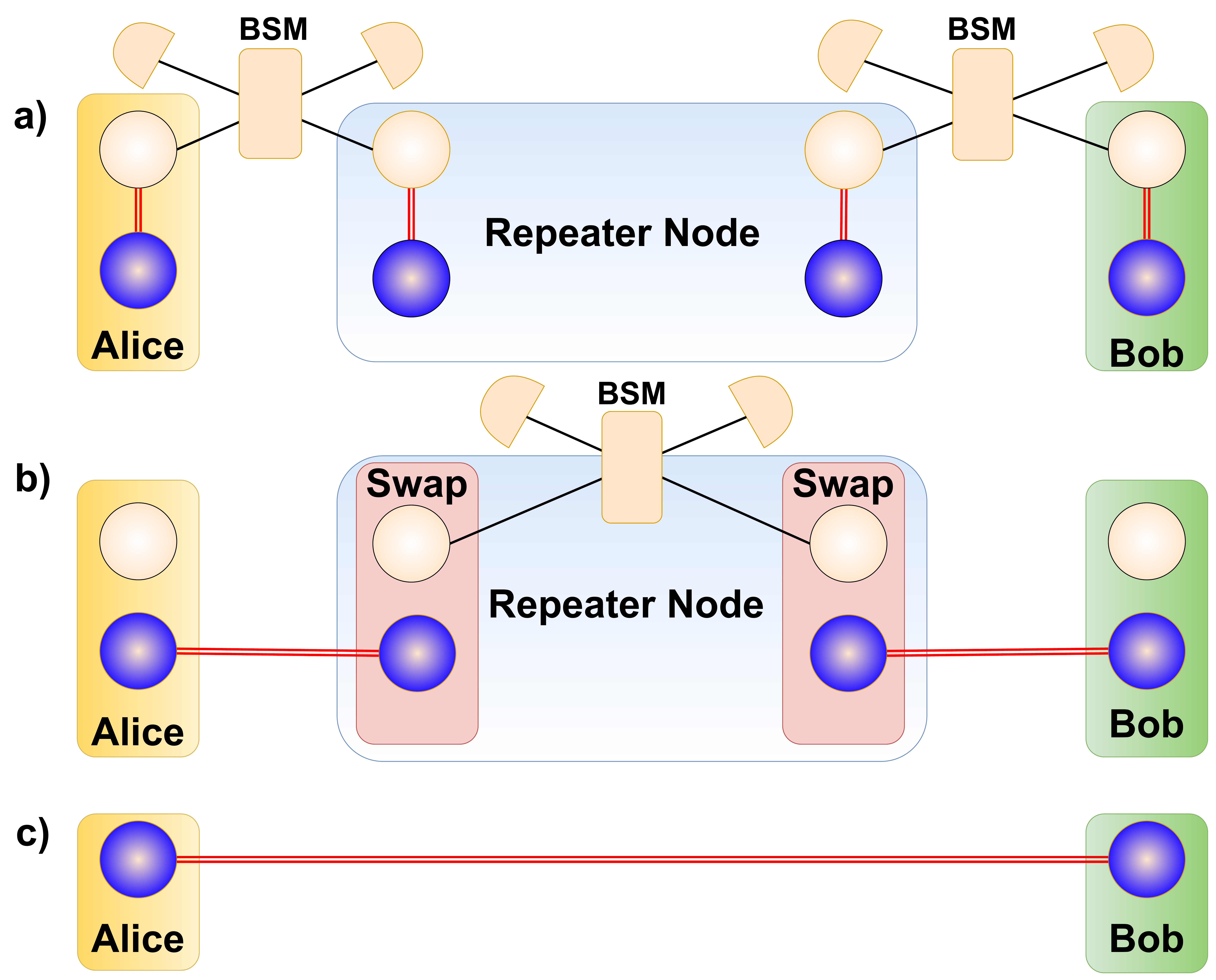}
    \caption{Illustration of a cavity magnon repeater network having cavities $c_1, c_2, c_2, c_4$ in pale yellow and magnons $m_1, m_2, m_2, m_4$ in blue circles from left to right. Doubled red lines represent entanglement between subsystems.}
    \label{rep}
\end{figure}

In Fig. (\ref{fig:singlet}), the overall purity of each Bell state exceeds ninety-four percent, as the expected basis populations reach about 0.48–0.49, while the matching off-diagonal coherence reaches $\pm$ 0.47, very close to the ideal mark of 0.5. This is due to the noise effect of $\approx$ (0.01–0.03) considered in the system. This confirms that our setup reliably creates and tells apart the four maximally entangled Bell states. The depolarization noise turns the perfect Bell pair into a $p=0.94$ Bell state, which will be further reduced along the length of the fiber optics channel. Furthermore, based on an extension of this methodology, we present a practical and scalable setup for implementing quantum repeaters based on hybrid cavity-magnon systems. The architecture is based on two-step entanglement swapping, i.e., one swap is over the noisy channel and another swap is locally at the repeater node as illustrated in Fig. (\ref{rep}). This architecture is designed to facilitate long-distance quantum communication by leveraging the strong coherence properties of magnons and high-quality microwave cavities with localized magnon modes found in ferromagnetic materials. This enables the robust generation, storage, and swapping of entanglement between distant nodes via fiber links, and subsequently extends it through entanglement swapping operations. We break the complete procedure into steps below,

\subsection{Preparation}
We consider four different nodes, and each node is initialized in an identical singlet Bell state as discussed above. %similar to Eq. \ref{eq3} but additional noise is taken into account. 
The mathematical expression for the density matrix is,
\begin{align*}
    \rho_{loc} = \bigotimes_{n=1}^{4} \rho_{c_{n}m_{n}}\
\end{align*}
where \(\rho_{c_{n}m_{n}}= p\ket{\psi^-}\bra{\psi^-}_n +(1- p)\rho_{noise}\) are the density matrix of the singlet bell generated between the cavity-magnon of each n$^{th}$ node. Here $p$ is the purity of the singlet Bell state, which is reduced due to the noise accumulating with length and each new interaction. Thus, leading to a mixed state with $1-p$ probability.

\subsection{First-Stage Entanglement Swapping}
Photons from cavities \(c_1\) and \(c_2\) and similarly from \(c_3\) and \(c_4\) interfere on a beam splitter and are subsequently detected via a BSM, as shown in Fig. \ref{rep}a. Each BSM projects the corresponding pair of magnons \((m_1,m_2)\) and \((m_3,m_4)\) onto one of the four Bell states. In practice, the particular Bell state obtained is random and determined by the BSM outcome. At this point, a classical feed-forward channel conveys the two-bit Bell outcome to the repeater node and the end node, and a conditional Pauli correction $\mathcal{U}_{cor_{i}}=\sigma_i$, where $i \in \{ x,y,z \}$, is applied so that in all cases the remaining two-magnon state rotates into the singlet as shown in Table (\ref{bsa}). For simplicity, we consider the case where both measurements project their magnon pairs onto the singlet state \(\ket{\psi^-}\bra{\psi^-}\). Under this assumption, no correction is needed, and the joint magnonic state at the repeater node becomes \(\rho \;=\; \rho_{m_1, m_2}\;\otimes\;\rho_{m_3, m_4}\quad\text{with}\quad\rho_{m_i m_j}=\ket{\psi^-}\bra{\psi^-}_{m_i, m_j}\,\), giving us adjacent magnon-magnon entanglement.

\subsection{Final SWAP Gate at Repeater Node}
At the repeater node, the local cavities \(c_2\) and \(c_3\) are initialized in their vacuum states \(\lvert0\rangle_{c_2}\lvert0\rangle_{c_3}\).  The quantum information stored in magnons \(m_2\) and \(m_3\) is then swapped into their respective cavities by the SWAP gate, which can be synthesized by another unitary evolution similar to Eq. (\ref{intHam}).

\begin{table}[H]
\centering
\caption{Bell‐state measurement projectors and feed‐forward Pauli corrections}
\label{bsa}
\begin{tabular}{c c c}
\toprule
 Outcome 
  & $\rho^{j} = \ket{\Phi_j}\bra{\Phi_j}$ 
  & $\mathcal{U}_{cor_i}$ \\ 
\midrule
$\displaystyle\ket{\psi^+} = \tfrac{1}{\sqrt2}(\ket{01}+\ket{10})$ 
  & $\rho^{0}$ 
  & $\sigma_z$ \\

$\displaystyle\ket{\psi^-} = \tfrac{1}{\sqrt2}(\ket{01}-\ket{10})$ 
  & $\rho^{1}$ 
  & $\mathbb{I}$ \\

$\displaystyle\ket{\phi^+} = \tfrac{1}{\sqrt2}(\ket{00}+\ket{11})$ 
  & $\rho^{2}$ 
  & $\sigma_z \sigma_x$ \\

$\displaystyle\ket{\phi^-} = \tfrac{1}{\sqrt2}(\ket{00}-\ket{11})$ 
  & $\rho^{3}$ 
  & $\sigma_x$ \\
\bottomrule
\end{tabular}
\end{table}
Furthermore, SWAP operation is performed between cavities \(c_2\) and \(c_3\).  A subsequent BSM on these two cavities then heralds the successful generation of end‐to‐end entanglement between the distant magnons \(m_1\) and \(m_4\) after applying the unitary correction as shown in Table (\ref{bsa}) for feed forward as discussed in \cite{hauser_boosted_2025}. This cancels any mismatch, mitigates any additional losses and adds to the total success probability of the system. Another factor that adds to the probability of success is the multiplexing $M$ of the entangled state in the fiber channel. We consider different levels of multiplexing in the fiber channel based on the experimentally relevant literature \cite{sinclair_spectral_2014,heinsoo_rapid_2018,kitayama_dense_2003,dhara_heralded_2022}. Finally, a reset signal brings all the cavities back to the initialized state and repeats the process for $n$ number of nodes. 

\section{Results and Discussion}\label{resultsNdiscussion}

To evaluate the practical feasibility of our cavity-magnon quantum repeater architecture, we perform extensive numerical simulations that account for realistic experimental conditions and system imperfections using the analytical expressions provided in Table (\ref{tab:formulas}) and the parameter sets detailed in Table (\ref{tab:scenario_params}). Our analysis evaluates end-to-end performance across multiple network hops, considering both the quality of distributed entanglement and the probability of successful operations. We quantify repeater performance using three key metrics, concurrence as a measure of entanglement strength between distant magnons, fidelity comparing the final distributed state to the ideal Bell state, and success probability accounting for all sources of loss and measurement failures throughout the protocol.

\begin{table}
\centering
\caption{Analytical expressions used in the repeater‐chain simulations.}
\label{tab:formulas}
\resizebox{\columnwidth}{!}{%
\begin{tabular}{@{}lll@{}}
\toprule
Symbol & Meaning & Formula \\ 
\midrule
$\eta_{\mathrm{link}}(L)$
& Single‐photon link efficiency 
& $10^{-\alpha L/10}\,\eta_{conv}^2\,\eta_{\mathrm{extra}}$ \\[6pt]

$\mathcal{P}_{\mathrm{click}}$  
& Raw Bell‐click probability (one channel) 
& $\mathcal{P}_{\mathrm{raw}}\,\eta_{\mathrm{BSA}}\,\eta_{\mathrm{det}}^2\,\eta_{\mathrm{col}}^2\,\eta_{\mathrm{link}}^2\,p_{\mathrm{scale}}$ \\[6pt]

$\mathcal{P}_{\mathrm{succ}}^{(\mathrm{hop})}$
& Per‐hop success with multiplexing
& $1-\bigl(1-\mathcal{P}_{\mathrm{click}}\bigr)^{M_{\mathrm{mux}}}$ \\[6pt]

$\mathcal P_{h}$  
& Cumulative success after $h$ hops 
& $\prod_{j=1}^h \mathcal{P}_{\mathrm{succ}}^{(j)}$ \\[6pt]

$\mathcal F$  
& Conditional magnon–magnon fidelity 
& $\langle\psi^-|\rho_{m_1m_{h+1}}|\psi^-\rangle$\\[6pt]

$T_{\mathrm{swap}}$  
& Beam‐splitter evolution time 
& $\pi/(2\,g_{\mathrm{bs}})$ \\
\bottomrule
\end{tabular}
}
\end{table}

\begin{table*}
  \caption{Parameter set used in the 4-hop cavity–magnon repeater study.
           Chip-scale attenuation $\alpha$ is given in \si{dB\,cm^{-1}}
           and span length $L_{\text{span}}$ in centimetres; metro-fibre
           values are in \si{dB\,km^{-1}} and kilometres, respectively.
           A dash “--” indicates that microwave–to–optical conversion is
           not present (purely microwave link) \cite{sangouard_quantum_2011}.}
  \label{tab:scenario_params}
  \centering
  \begin{tabular}{lccccccccc}
    \toprule
    \textbf{Scenario} &
      $\alpha$ &
      $L_{\text{span}}$ &
      $\eta_{\text{read}}$ &
      $\eta_{\text{conv}}$ &
      loss$_{\text{extra}}$ &
      $\eta_{\text{det}}$ &
      $\eta_{\text{col}}$ &
      $P_{\text{BSA}}$ &
      $M_{\text{mux}}$ \\[2pt]
    \midrule
    Chip-A (1 cm)                     & 0.20\,dB\,cm$^{-1}$ & 1\,cm   & 0.62 & --    & 0.98 & 0.98 & 0.95 & 0.50 & 1  \\
    Chip-B (1 cm, multiplexed)             & 0.20\,dB\,cm$^{-1}$ & 1\,cm   & 0.62 & --    & 0.98 & 0.98 & 0.95 & 0.50 & 8  \\
    Chip-C (1 cm, ancilla, dense multiplexed)   & 0.20\,dB\,cm$^{-1}$ & 1\,cm   & 0.62 & --    & 0.98 & 0.98 & 0.95 & 0.75 & 30 \\[2pt]
    Metro-A (10 km)            & 0.35\,dB\,km$^{-1}$ & 10\,km  & 0.62 & 0.005 & 0.90 & 0.80 & 0.95 & 0.50 & 1  \\
    Metro-B (10 km, multiplexed)      & 0.20\,dB\,km$^{-1}$ & 10\,km  & 0.62 & 0.50  & 0.95 & 0.98 & 0.95 & 0.50 & 8  \\
    Metro-C (10 km, dense multiplexed)           & 0.16\,dB\,km$^{-1}$ & 10\,km  & 0.62 & 0.80  & 0.95 & 0.98 & 0.95 & 0.75 & 30 \\
    \bottomrule
  \end{tabular}
\end{table*}
\begin{figure*}[!t]
  \centering
  \includegraphics[width=\textwidth]{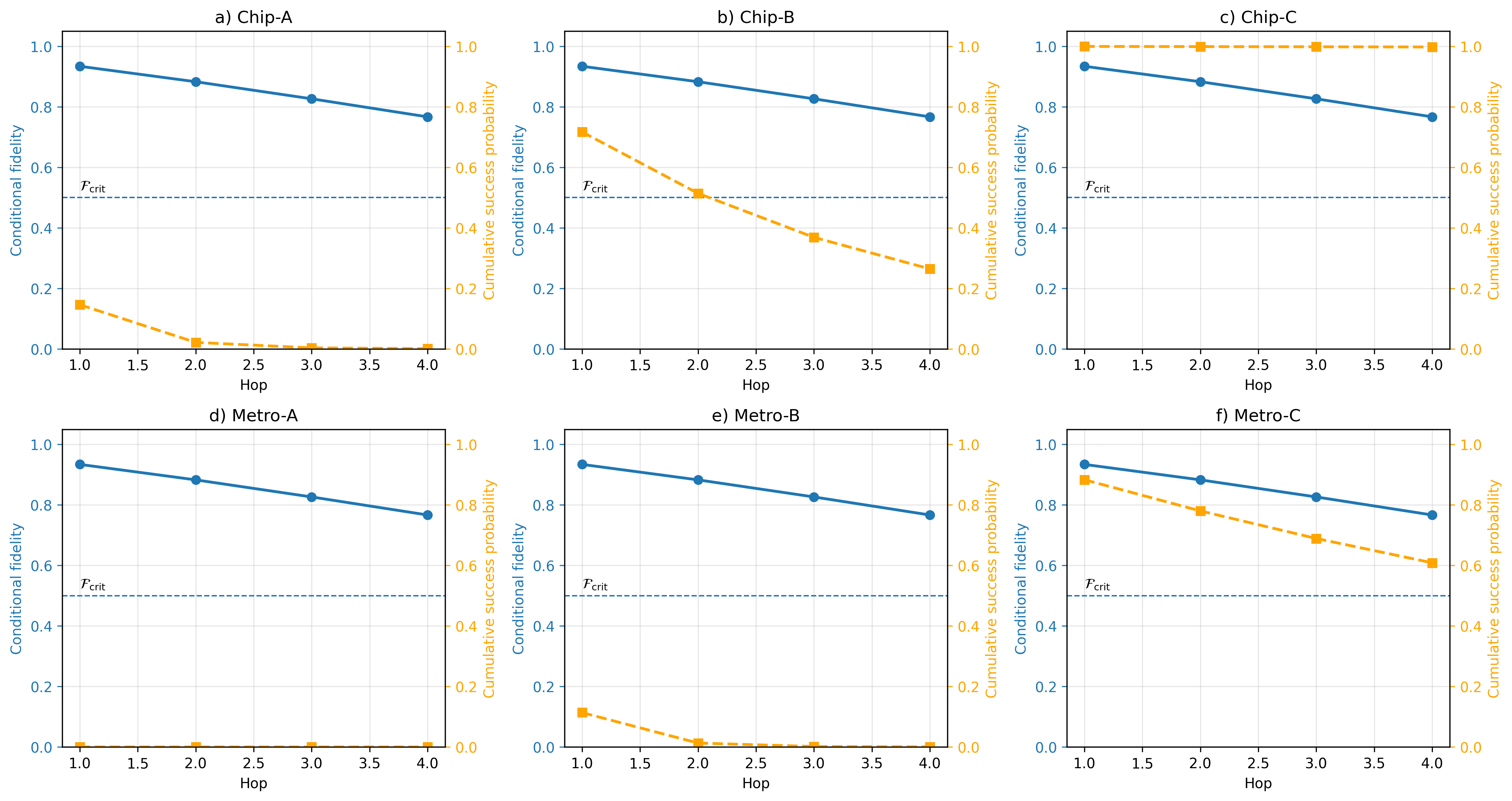}
  \caption{a) On-Chip (1 cm), b) Metro-scale (10 km), c) Metro-scale (10 km). In each plot, the solid blue line represents the fidelity of the distributed state with respect to the initial Bell state $\psi^-$. The blue dashed line indicates the fidelity threshold beyond which entanglement is effectively lost, rendering the network operationally useless for quantum communication. The yellow dashed line corresponds to the cumulative success probability after each hop in the entanglement distribution process. }
  \label{hopnet}
\end{figure*}

Our simulations model a four-hop quantum repeater chain, where each hop represents the distance between adjacent network nodes. We consider two distinct deployment scenarios: chip-scale implementation with 1 cm links operating entirely in the microwave domain, and metro-scale deployment with 10 km fiber links requiring microwave-to-optical conversion. The total success probability incorporates multiple loss mechanisms, where for a single photon traversing distance $L$, the link efficiency follows $\eta_{link} = 10^{-\alpha L /10} \eta^2_{conv}\eta_{extra}$, with $\alpha$ representing fibre attenuation, $\eta_{conv}$ accounting for microwave-to-optical conversion efficiency, and $\eta_{extra}$ capturing parasitic losses. The raw Bell-state measurement probability becomes $\mathcal{P}_{click} = \mathcal{P}_{\mathrm{raw}}\,\eta_{\mathrm{BSA}}\,\eta_{\mathrm{det}}^2\,\eta_{\mathrm{col}}^2\,\eta_{\mathrm{link}}^2\,p_{\mathrm{scale}}$, explicitly factoring in Bell-state analyzer efficiency, detector efficiency, collection efficiency, and wavelength-dependent penalties.

Fig. (\ref{hopnet}a) presents the performance characteristics for chip-scale deployment with 1 cm links. Starting from an initial Bell state with 0.97 concurrence, our simulations demonstrate robust entanglement preservation across four network hops. The fidelity exhibits a gradual decline from 0.95 after the first swap to approximately 0.78 after the fourth hop, remaining well above the 0.7 threshold required for practical quantum communication applications. However, the success probability follows a more dramatic progression, dropping from an initial 18\% for the first hop to below 5\% after four hops when operating without multiplexing. This steep decline primarily results from the cumulative effect of Bell-state measurement inefficiencies, where each 50\% success probability compounds across multiple operations. 

The introduction of 8-channel multiplexing dramatically improves performance characteristics, as demonstrated in Fig. (\ref{hopnet}b). The per-hop success probability increases to approximately 80\%, showing how parallel operation across multiple spectral channels compensates for individual channel failures. The multiplexed success probability follows $\mathcal{P}_{\mathrm{succ}}^{(\mathrm{hop})} = 1-\bigl(1-\mathcal{P}_{\mathrm{click}}\bigr)^{M_{\mathrm{mux}}}$, where \(M_{\mathrm{mux}}\) represents the number of multiplexed channels. Fig. (\ref{hopnet}c) illustrates the performance under dense multiplexing conditions with 30 channels and advanced Bell-state analyzers achieving 75\% efficiency. Under these conditions, the cumulative success probability approaches unity while maintaining high fidelity performance, representing the theoretical upper bound for our architecture and demonstrating the scalability potential when combined with advanced detection techniques.

Metro-scale deployment introduces additional complexity through the requirement for microwave-to-optical conversion. Fig. (\ref{hopnet}d) shows the performance with realistic conversion efficiency $\eta_{conv}=0.005$ , representative of current experimental capabilities. The dramatically reduced success probability highlights the critical bottleneck imposed by inefficient conversion processes. However, Fig. (\ref{hopnet}e-\ref{hopnet}f) demonstrate the impact of improved conversion technologies. With moderate efficiency improvements ($\eta_{conv}=0.5$), the system recovers substantial functionality, achieving success probabilities comparable to chip-scale implementations under multiplexing conditions. The highest performance scenario ($\eta_{conv}=0.8$) approaches the theoretical limits, suggesting that advances in microwave-to-optical interfaces will be crucial for practical metro-scale deployments.

The 10 km fiber links introduce additional photon loss through standard telecom fiber attenuation ($\alpha$ = 0.16-0.35 dB/km). Our simulations show that this distance-dependent loss, while significant, can be compensated through multiplexing and improved detector efficiency, particularly when combined with high-efficiency conversion technologies. The relationship between multiplexing level and system performance follows predictable scaling laws, where for $M_{mux}$ multiplexed channels, the effective success probability scales as $1 - (1 - \mathcal{P}_{click})^M_{mux}$, with $\mathcal{P}$ representing the single-channel success probability. This scaling provides diminishing returns, the improvement from 1 to 8 channels is substantial (18\% - 78\%), while the gain from 8 to 30 channels is more modest (78\% - 98\%).

Dense multiplexing imposes practical constraints on system complexity, requiring sophisticated wavelength division multiplexing infrastructure and high-bandwidth detection systems for the 30-channel implementation \cite{yoshida_multiplexed_2024}, \cite{kitayama_dense_2003, heinsoo_rapid_2018}. However, our analysis suggests that moderate multiplexing (8 channels) captures most of the performance benefits while maintaining reasonable implementation complexity. While our analysis focuses on spectral multiplexing through wavelength division, the framework extends naturally to temporal multiplexing schemes, with recent experiments demonstrating both approaches where spectral multiplexing offers advantages for continuous operation and temporal multiplexing provides simpler hardware requirements.

Quantum communication protocols typically require fidelity greater than 0.7 for secure key distribution applications, and our architecture comfortably exceeds this threshold across all scenarios, with the lowest fidelity (0.78) occurring in the four-hop metro-scale configuration  \cite{munro_quantum_2012, chehimi_scaling_2023}. This margin provides robustness against additional sources of error not captured in our idealized simulations. Compared to atomic ensemble quantum memories, our cavity-magnon approach offers superior integration capabilities and longer coherence times, while relative to solid-state platforms like nitrogen-vacancy centers, magnonic systems provide better frequency tunability and reduced sensitivity to magnetic field fluctuations. The hybrid nature of our platform enables interfaces with both superconducting and optical quantum systems.

The heralded success probabilities translate directly to communication throughput, where under optimal conditions (dense multiplexing, advanced BSA), our architecture supports kilohertz-level secure key generation rates, comparable to demonstrated quantum repeater systems while offering superior scalability prospects. Our parameter sets derive directly from state-of-the-art experimental demonstrations, ensuring that the simulated performance reflects achievable hardware capabilities. The YIG sphere parameters (Q-factors \(10^4-10^6\), coupling rates 130 MHz) represent current experimental standards, while the projected improvements in conversion efficiency align with ongoing research trajectories.

The modular architecture of our repeater design enables straightforward network expansion, where additional hops follow the same scaling laws with performance degrading predictably according to the cumulative loss model. This predictability facilitates network planning and resource allocation for large-scale deployments. Real implementations must address several practical challenges not captured in our idealized simulations, including frequency stabilization across network nodes, synchronization of entanglement swapping operations, and integration with classical control systems for feed-forward operations. However, the magnonic platform's inherent stability and controllability provide advantages for addressing these challenges.

Our analysis employs several simplifying assumptions that allow tractable modeling of the system. The Lindblad master equation approach provides a phenomenological treatment of decoherence but may not capture all environmental effects, and the assumption of perfect spatial mode matching between cavity and magnon modes represents an idealization that requires careful experimental optimization. Several technical hurdles must be overcome for practical implementation, particularly the development of high-efficiency, low-noise microwave-to-optical converters, which remains a critical bottleneck for metro-scale deployment. Additionally, the realization of high-fidelity Bell-state analyzers with deterministic outcomes would dramatically improve system performance.

\section{Conclusion}\label{conclusion}
In this paper, we developed and analyzed a comprehensive cavity-magnon quantum repeater architecture for scalable quantum communication networks. Through extensive numerical simulations incorporating realistic experimental parameters, we demonstrated the viability of these hybrid systems for both chip-scale and metro-scale deployments. Our cavity-magnon platform offers unique advantages, including excellent frequency tunability, reduced environmental sensitivity, and seamless integration with both superconducting and optical quantum systems, positioning it as an ideal interface for heterogeneous quantum networks.
Our analysis reveals that while chip-scale implementations achieve near-optimal performance, metro-scale deployment faces critical bottlenecks in microwave-to-optical conversion efficiency. The development of high-efficiency, low-noise converters is a primary technological challenge that must be addressed for practical long-distance quantum networks. Additionally, practical implementation requires advances in frequency stabilization across distributed nodes, synchronization of entanglement swapping operations, and integration with classical control systems for real-time network management.

The scalability of our architecture, through spectral multiplexing, provides a clear pathway for performance enhancement while maintaining reasonable implementation complexity. Future research should prioritize conversion technology development through electro-optic and optomechanical approaches, exploration of hybrid multiplexing schemes, and investigation of alternative network topologies beyond linear chains. Integration with quantum error correction protocols represents another crucial direction for enabling fault-tolerant distributed quantum computation.

Our cavity-magnon quantum repeater architecture establishes a technologically mature foundation for next-generation quantum networks. Its compatibility with existing fiber-optic infrastructure and demonstrated performance benchmarks position these systems as a viable bridge between laboratory demonstrations and practical quantum communication applications. As quantum information technologies transition toward real-world deployment, our work provides both the theoretical framework and practical roadmap for implementing scalable quantum repeater networks that can support the emerging quantum internet infrastructure.
%%%%%%%%%%%%%%%%%%%%%%%%%%%%%%%%%%%%%%%%%%%%%%%%%%%%%%%%%%%%%%%%
\begin{center}
    {\Large \textbf{Declarations}}
\end{center}
\begin{flushleft}
\textbf{Ethical Approval }
\end{flushleft}
We hereby, undersigned declare that this manuscript contains original work and it has not been published before or
currently being considered for publication elsewhere. Appropriate references have been made for mention of information from else work.
\begin{flushleft}
\textbf{Competing interests}
\end{flushleft}
 We confirm that there are no known conflict of interests associated with this publication and there is no conflict by any of the authors that have or can effect the result and time of publication.
\begin{flushleft}
\textbf{Authors' contributions}
\end{flushleft} 
The manuscript has been read and approved by all named authors and there are no other persons who satisfies the criteria for authorship but is not listed. We confirm that all have approved the order of authors listed in the named authors' manuscripts. We confirm that we have given due consideration to the protection of intellectual property associated with this work and that there are no obstruction to publication.
\begin{flushleft}
\textbf{Funding}
\end{flushleft} 
 Will fill this section after discussion
\begin{flushleft}
\textbf{Availability of data and materials}
\end{flushleft} 
The material presented in the document is owned by the authors, and/no permissions are required.

\bibliographystyle{ieeetr}
\bibliography{Refs}

\end{document}